\newcommand{\diracslash}[1]{#1\llap{/\kern2pt}}

\newcommand{\be}{\begin{equation}}
\newcommand{\ee}{\end{equation}}
\newcommand{\bea}{\begin{eqnarray}}
\newcommand{\eea}{\end{eqnarray}}
\newcommand{\ba}[1]{\begin{array}{#1}}
\newcommand{\ea}{\end{array}}

\documentclass[prd,aps,floats,nofootinbib,tightenlines,showpacs]{revtex4}
\usepackage{epsfig,graphicx,pstricks}
\usepackage{psfrag}
\usepackage{color}
\usepackage{amsmath}
\usepackage{amsfonts}
\usepackage{amssymb}
\usepackage{textcomp}
\usepackage{multirow}
\usepackage{natbib}
\usepackage{subfigure}
\begin{document}

\title { Higher moments on strangeness fluctuation using PNJL model }
\author{Paramita Deb }
\email{paramita.dab83@gmail.com}
\affiliation{Department of Physics, Indian Institute of Technology Bombay,
Powai, Mumbai- 400076, India}
\author{Amal Sarkar} 
\email{amal@rcf.rhic.bnl.gov}
\affiliation{Department of Physics, NRF iThemb LABS,
Cape Town, South Africa}
\author{Raghava Varma}
\email{varma@phy.iitb.ac.in}
\affiliation{Department of Physics, Indian Institute of Technology Bombay
Powai, Mumbai-400076, India}

\date{\today} 

\def\be{\begin{equation}}
\def\ee{\end{equation}}
\def\bearr{\begin{eqnarray}}
\def\eearr{\end{eqnarray}}
\def\zbf#1{{\bf {#1}}}
\def\bfm#1{\mbox{\boldmath $#1$}}
\def\hf{\frac{1}{2}}
\def\sl{\hspace{-0.15cm}/}
\def\omit#1{_{\!\rlap{$\scriptscriptstyle \backslash$}
{\scriptscriptstyle #1}}}
\def\vec#1{\mathchoice
        {\mbox{\boldmath $#1$}}
        {\mbox{\boldmath $#1$}}
        {\mbox{\boldmath $\scriptstyle #1$}}
        {\mbox{\boldmath $\scriptscriptstyle #1$}}
}

\begin{abstract}

\end{abstract}


\maketitle

\section{Introduction}
The strongly interacting matter is supposed to have a rich phase structure at 
finite temperature and density. While our Universe at present epoch contains a
significant fraction of color singlet hadrons,
color non-singlet states especially quarks and gluons may have been prevalent 
in the few microseconds after the Big bang. One of the fundamental goals of the 
heavy ion collision experiments is to map the QCD phase diagram and to locate 
the critical end point (CEP), where the first order phase transition from hadronic state to
quark gluon plasma (QGP) phase becomes continuous {\cite{STAR, STAR1}}. Presently, 
neither the existence nor the exact location of the critical point is known in spite of the
heavy ion collision experiments being carried out at Relativistic Heavy
Ion Collider (RHIC) at BNL and Super Proton Synchroton (SPS) at CERN. Further, it is 
equally important  to choose the correct experimental observables that will help locate the critical 
point. In the heavy ion collision experiments, formation of QGP is followed by expansion and
attainment of freeze out characterized by cessation of all chemical and kinetic interactions. 
The detected particles at the freeze out condition can lead to the location of the freeze out point. 
Thus to locate the critical point, experiments are being conducted to bring the freeze out point  
close to the critical point by varying the collision center of mass energy
$\sqrt s$. Therefore, there is a need to select the suitable experimental observables such as fluctuations 
of the conserved quantities that are sensitive to the 
proximity of the freeze-out point. The fluctuations of an experimental observable is defined as the 
variance and higher non-Gaussian moments of the event-by-event 
distribution for an experimental observable of each event in an ensemble of many events. 
These fluctuations result in a long range correlation length 
$\xi; \text{maximal value} \approx 1.5 - 3 fm$ \cite {rajagopal2000}. 
Hence the non-monotonic behavior of these fluctuations could be the signature of the critical
point \cite{rajagopal1998}. As different particles correspond to  different conserved quantum numbers 
like baryon number $(B)$, electric charge $(Q)$ and strangeness $(S)$; an
event-by-event analysis of fluctuation of these can help locate the 
critical end point. 

The QGP matter formed in the heavy ion collision experiments has a 
finite volume 
depending on the size of the colliding nuclei, center of mass energy and
collision centrality. Several efforts have been made to estimate the finite
volume during the freeze-out for different centrality measurement of HBT radii 
{\cite {adamova}}. These results suggest that the volume increases with the centrality during 
freeze out and it is estimated to be $2000 fm^3$ to $3000 fm^3$. Theoretically 
the effects of finite volume have been addressed by
many models such as non-interacting bag model {\cite {elze}}, 
chiral perturbation theory {\cite {luscher,gasser}}, 
Nambu--Jona-Lasinio (NJL) model {\cite {nambu,kiriyama,shao}}, 
linear sigma model {\cite {braun,braun1}} 
 and  by the first principle study of pure gluon theory on space time 
 lattices {\cite {gopie,bazavov}}. Specifically, in a $1+1$ dimensional NJL model the 
finite size effect of a dense baryonic matter has been described by the induction
of a charged pion condensation phenomena. Recently, this has been extended to Polyakov 
loop Nambu--Jona-Lasinio (PNJL) model where it was observed that as the 
volume decreases, critical temperature for the 
crossover transition decreases. For lower volumes, CEP is shifted to a domain with  higher 
chemical potential $(\mu)$ and lower temperature $(T)$ {\cite {deb1,abhijit}.
It is quite evident that broadly both the Lattice calculations and QCD-based models
indicate that the fluctuation of strongly interacting matter
at zero density show significant volume dependence which might be relevant to
study the formation of fireball in heavy-ion collision. 

Further, both the Lattice QCD 
results {\cite{boyd,engels,fodor,allton,forcrand,aoki,megias}} and the QCD 
inspired models {\cite {fukushima,ratti,pisarski,fukushima1,hansen,ciminale,
ghosh,deb,deb1,osipov,kashiwa,schaefer, mustafa}} show that
the net conserved quantum numbers ($B$, $Q$ and $S$) are related to the conserved number susceptibilities 
($\chi_x= \langle(\delta N_x)^2\rangle/VT $ where $x$ can be either $B$, $S$ or $Q$ 
and $V$ is the volume).
Close to the critical point,
models also predict that the distributions of the 
conserved quantum numbers to be non-Gaussian and 
susceptibilities  to diverge causing both skewness $(s)$ and kurtosis 
$(\kappa)$ to deviate where 
$s\sigma = (\chi_x^3/\chi_x^2)$ and $\kappa\sigma^2 = \chi_x^4 / \chi_x^2$. These quantities  
are much more sensitive (skewness $\sim \xi^{4.5}$ and kurtosis $\sim \xi ^7$) 
to the correlation length and they can provide much better handle for location 
of CEP. Moreover, the higher order coefficients become increasingly sensitive 
in the vicinity of phase transition.
For example, in a 2 flavor QCD model it has been shown that the baryon number fluctuation $(\chi_B)$ increases
with temperature  and its fourth moments attains a maxima
in the phase transition region from low to high temperature {\cite{ejiri}}.  
Similarly, fluctuations have been also computed with respect to 
the quark chemical potential $(\mu)$ in the 
Polyakov loop coupled quark-meson (PQM) model {\cite{schaefer}} and its 
renormalized group improved version, 2 flavor PNJL model with three-momentum 
cutoff regularization {\cite{roessner}}. Fluctuations and the correlations of conserved 
charges have also been studied in higher flavor PNJL model \cite {deb,deb1, abhijit, upadhaya} with or 
without finite volume effects as well as simplistic lattice QCD  {\cite{cheng}}. Recently, a realistic
continuum limit calculation \cite {bazavov1,bazavov2,borsanyi} for the lattice 
QCD data has been performed and 
the 3 flavor PNJL model parameters have been reconsidered \cite{saha}. 
This re-parametrization has indeed resulted in a very good quantitative agreement
between the model and the lattice data at finite temperature and zero 
density region. The second order and fourth order susceptibilities of the 
baryon number were found to be in reasonable quantitative agreement with the lattice data. 
For electric charge susceptibilities there were some disagreement for the temperature less than the critical temperature. 
However, in order to understand the
QCD phase diagram and find the critical end point one need to explore the finite density
region. Current work will emphasis on the 3 flavor finite volume finite density 
PNJL model to study the strangeness susceptibilities $(\chi_S)$ and 
correlation among different conserved charges and compare them to both the 
recent experimental finding
and the hadron resonance gas model (HRG) {\cite{braun2001, 
andronic2009a}}  data which is generally considered as a theoretical baseline for comparing
the experimental results as well as other theoretical models.

In the context of the above discussions, we organize the present work as follows. We describe 
the thermodynamic formulation of the 3-flavor finite volume PNJL model with six-quark and 
eight-quark interactions. Subsequently,  the method 
to calculate the correlations of conserved charges in PNJL model has been elaborated. Finally, the 
variation of skewness $(s)$, kurtosis $(\kappa)$ of strangeness fluctuations 
$(\chi_S)$ and higher moments of cross-correlations with collision energy has 
been determined.

\section{The PNJL model
}
We shall consider the 2+1 flavor PNJL model with six quark and eight quark 
interactions. In the PNJL model
the gluon dynamics is described by the chiral point couplings
between quarks (present in the NJL part) and a background gauge field 
representing Polyakov Loop dynamics. The Polyakov line is represented as,
\begin {equation}
  L(\bar x)={\cal P} {\rm exp}[i {\int_0}^\beta
d\tau A_4{({\bar x},\tau)}]
\end {equation}
where $A_4=iA_0$ is the temporal component of Eucledian gauge field
$(\bar A,A_4)$, $\beta=\frac {1}{T} $, and $\cal P$ denotes path
ordering. $L(\bar x)$ transforms as a field with charge one under
global Z(3) symmetry. The Polyakov loop is then given by 
$\Phi = (Tr_c L)/N_c$, and its conjugate by,
${\bar \Phi} = (Tr_c L^\dagger)/N_c$. The gluon dynamics can be
described as an effective theory of the Polyakov loops. Consequently,
the Polyakov loop potential can be expressed as,
\begin{equation}
\frac {{\cal {U^\prime}}(\Phi[A],\bar \Phi[A],T)} {T^4}= 
\frac  {{\cal U}(\Phi[A],\bar \Phi[A],T)}{ {T^4}}-
                                     \kappa \ln(J[\Phi,{\bar \Phi}])
\label {uprime}
\end{equation}
where $\cal {U(\phi)}$ is a Landau-Ginzburg type potential commensurate
with the Z(3) global symmetry. Here we choose a form given in
\cite{ratti},
\begin{equation}
\frac  {{\cal U}(\Phi, \bar \Phi, T)}{  {T^4}}=-\frac {{b_2}(T)}{ 2}
                 {\bar \Phi}\Phi-\frac {b_3}{ 6}(\Phi^3 + \bar \Phi^3)
                 +\frac {b_4}{  4}{(\bar\Phi \Phi)}^2,
\end{equation}
where
\begin {eqnarray}
     {b_2}(T)=a_0+{a_1}exp(-a2{\frac {T}{T_0}}){\frac {T_0}{T}}
\end {eqnarray}
$b_3$ and $b_4$ being constants. The second term in eqn.(\ref {uprime})
is the Vandermonde term which replicates the effect of SU(3) Haar
measure and is given by,
\begin {equation}
J[\Phi, {\bar \Phi}]=(27/24{\pi^2})\left[1-6\Phi {\bar \Phi}+\nonumber\\
4(\Phi^3+{\bar \Phi}^3)-3{(\Phi {\bar \Phi})}^2\right]
\end{equation}
The corresponding parameters were earlier obtained in the above
mentioned literature by choosing suitable values by fitting a few
physical quantities as function of temperature obtained in LQCD
computations. The set of values chosen here are listed in the 
table \ref{table1} \cite{saha}. 
\vskip 0.1in
\begin{table}[htb]
\begin{center}
\begin{tabular}{|c|c|c|c|c|c|c|c|c|c|c|c|}
\hline
Interaction & $ T_0 (MeV) $ & $ a_0 $ & $ a_1 $ & $ a_2 $ & $ b_3 $ &$
b_4$ & $  \kappa $ \\ 

\hline
6-quark &$ 175 $&$ 6.75 $&$ -9.0 $&$ 0.25 $&$ 0.805 $&$7.555 $&$ 0.1 $ \\

\hline

8-quark & $ 175 $&$ 6.75 $&$ -9.8 $&$ 0.26 $&$0.805$&$ 7.555 $&$ 0.1 $\\
\hline

\end{tabular}
\caption{Parameters for the Polyakov loop potential of the model.}  
\label{table1}
\end{center}
\end{table}
\vskip 0.1in

 For the quarks we shall use the usual form of the NJL model except
for the substitution of a covariant derivative containing a background
temporal gauge field. Thus the 2+1 flavor the Lagrangian may be written as,
\begin{equation}
\begin{split}
   {\cal L} = {\sum_{f=u,d,s}}{\bar\psi_f}\gamma_\mu iD^\mu
             {\psi_f}&-\sum_f m_{f}{\bar\psi_f}{\psi_f}
              +\sum_f \mu_f \gamma_0{\bar \psi_f}{\psi_f}
       +{{g_S}\over{2}} {\sum_{a=0,\ldots,8}}[({\bar\psi} \lambda^a
        {\psi})^2+
            ({\bar\psi} i\gamma_5\lambda^a {\psi})^2] \nonumber\\
       &-{g_D} [det{\bar\psi_f}{P_L}{\psi_{f^\prime}}+det{\bar\psi_f}
            {P_R}{\psi_{f^\prime}}]\nonumber\\
 &+8{g_1}[({\bar\psi_i}{P_R}{\psi_m})({\bar\psi_m}{P_L}{\psi_i}]^2
    +16{g_2}[({\bar\psi_i}{P_R}{\psi_m})({\bar\psi_m}{P_L}{\psi_j})
          ({\bar\psi_j}{P_R}{\psi_k})({\bar\psi_k}{P_L}{\psi_i})]\nonumber\\
                 &-{\cal {U^\prime}}(\Phi[A],\bar \Phi[A],T)
%\label{lag2}
\end{split}
\end{equation}
where $f$ denotes the flavors $u$, $d$ or $s$ respectively.
The matrices $P_{L,R}=(1\pm \gamma_5)/2$ are respectively the
left-handed and right-handed chiral projectors, and the other terms
have their usual meaning, described in details in
Refs.~\cite{ghosh,deb,deb1}. This NJL part of the theory
is analogous to the BCS theory of superconductor, where the
pairing of two electrons leads to the condensation causing a gap in
the energy spectrum. Similarly in the chiral limit, NJL model exhibits
dynamical breaking of ${SU(N_f)}_L \times {SU(N_f)_R}$ symmetry to
$SU(N_f)_V$ symmetry ($N_f$ being the number of flavors). As a result 
the composite operators ${\bar \psi_f}\psi_f$ generate nonzero vacuum
expectation values. The quark condensate is given as,
\begin {equation}
 \langle{\bar \psi_f}{\psi_f}\rangle= 
-i{N_c}{{{\cal L}t}_{y\rightarrow x^+}}(tr {S_f}(x-y)),
\end {equation}
where trace is over color and spin states. The self-consistent gap 
equation for the constituent quark masses are,
\begin {equation}
  M_f =m_f-g_S \sigma_f+g_D \sigma_{f+1}\sigma_{f+2}-2g_1 
     \sigma_f{(\sigma_u^2+\sigma_d^2+\sigma_s^2)}-4g_2 \sigma_f^3  
\end {equation}
where $\sigma_f=\langle{\bar \psi_f} \psi_f\rangle$ denotes chiral 
condensate of the quark with flavor $f$. Here if we consider
$\sigma_f=\sigma_u$, then $\sigma_{f+1}=\sigma_d$ and
$\sigma_{f+2}=\sigma_s$, 
%$\sigma_{f+2}=\sigma_d$. 
The expression for $\sigma_f$ at zero
temperature ($T=0$) and chemical potential ($\mu_f=0$) may be
written as \cite{deb},
\begin {equation}
 \sigma_f=-\frac {3{M_f}}{ {\pi}^2} {{\int}^\Lambda}\frac {p^2}{
           \sqrt {p^2+{M_f}^2}}dp,
\end {equation}
$\Lambda$ being the three-momentum cut-off. This cut-off have been used
to regulate the model because it contains couplings with finite dimensions
which leads to the model to be non-renormalizable.

Due to the dynamical breaking of chiral symmetry, $N_f^2 - 1$ Goldstone
bosons appear. These Goldstone bosons are the pions and kaons whose masses, 
decay widths from experimental observations are utilized to fix the NJL model
parameters. The parameter values have been listed in table \ref{table2}.
Here we consider the $\Phi$, $\bar \Phi$ and $\sigma_f$ fields in the
mean field approximation (MFA) where the mean field are obtained by
simultaneously solving the respective saddle point equations.

\begin{table}[htb]
\begin{center}
\begin{tabular}{|c|c|c|c|c|c|c|c|c|c|c|}
\hline
Model & $ m_u (MeV) $ & $ m_s (MeV)$ & $ \Lambda (MeV) $ & $ g_S \Lambda^2 $ & $ g_D \Lambda^5 $ 
&$
g_1 \times 10^{-21} (MeV^{-8})$ & $ g_2 \times 10^{-22} (MeV^{-8})$ \\ 

\hline
With 6-quark &$ 5.5 $&$ 134.76 $&$ 631 $&$ 3.67 $&$ 9.33 $&$0.0 $&$ 0.0 $ \\

\hline

With 8-quark & $ 5.5 $&$ 183.468 $&$ 637.720 $&$ 2.914 $&$ 75.968
$&$ 2.193 $&$ -5.890 $ \\
\hline

\end{tabular}
\caption{Parameters of the Fermionic part of the model.}  
\label{table2}
\end{center}
\end{table}

 Now that the PNJL model is described for infinite volumes we discuss
how we implement the finite volume constraints. Ideally one should
choose the proper boundary conditions $-$ periodic for bosons and
anti-periodic for fermions. This would lead to a infinite sum over
discrete momentum values $p_i=\pi n_i/R$, where $i=x,y,z$ and $n_i$
are all positive integers and $R$ is the lateral size of the finite 
volume system. This implies a lower momentum cut-off
$p_{min}=\pi/R=\lambda$. One should also incorporate proper
effects of surface and curvatures. In this first case study we shall
however take up a number of simplifications listed below:

\begin{itemize}
\item
Surface and curvature effects have been neglected.
\item
The infinite sum will be considered as an integration over a continuous
variation of momentum albeit with the lower cut-off.
\item
Any modifications to the mean-field parameters due to
finite size effects will not be considered. Thus the Polyakov loop potential as well as the
mean-field part of the NJL model would remain unchanged. 
\end{itemize}

The thermodynamic potential for the multi-fermion interaction in MFA of the
PNJL model can be written as,

\begin {eqnarray}
 \Omega &=& {\cal {U^\prime}}[\Phi,\bar \Phi,T]+2{g_S}{\sum_{f=u,d,s}}
            {{\sigma_f}^2}-{{g_D} \over 2}{\sigma_u}
          {\sigma_d}{\sigma_s}+3{{g_1}\over 2}({{\sigma_f}^2})^2
           +3{g_2}{{\sigma_f}^4}-6{\sum_f}{\int_{0}^{\Lambda}}
     {{d^3p}\over{(2\pi)}^3} E_{pf}\Theta {(\Lambda-{ |\vec p|})}\nonumber \\
       &-&2{\sum_f}T{\int_0^\infty}{{d^3p}\over{(2\pi)}^3}
       [\ln\left[1+3(\Phi+{\bar \Phi}e^{-{(E_{pf}-\mu)\over T}})
       e^{-{(E_{pf}-\mu)\over T}}+e^{-3{(E_{pf}-\mu)\over T}}\right]\nonumber\\
       & +& \ln\left[1+3({\bar \Phi}+{ \Phi}e^{-{(E_{pf}+\mu)\over T}})
            e^{-{(E_{pf}+\mu)\over T}}+e^{-3{(E_{pf}+\mu)\over T}}\right]]
\end {eqnarray}
where $E_{pf}=\sqrt {p^2+M^2_f}$ is the single quasi-particle energy,
$\sigma_f^2=(\sigma_u^2+\sigma_d^2+\sigma_s^2)$ and 
$\sigma_f^4=(\sigma_u^4+\sigma_d^4+\sigma_s^4)$.
In the above integrals, the vacuum integral
has a cutoff $\Lambda$ whereas the medium dependent integrals
have been extended to infinity. The eight quark interaction in the Lagrangian
stabilize the vacuum. In the present study we have considered PNJL model with 
6-quark and 8-quark interactions for two sets of finite volume system with
lateral size $R=2 fm$ and $R=4 fm$. Thus we have four sets of parameter sets
(a) PNJL-6-quark for $R=2 fm$, (b) PNJL-6-quark for $R=4 fm$, (c) PNJL-8-quark 
for $R=2 fm$ and (d) PNJL-8-quark for $R=4 fm$. 

\vskip 0.2in
{\subsection{Taylor expansion of pressure}}
The freeze-out curve $T(\mu_B)$ in the $T-\mu_B$ plane and the dependence of 
the baryon chemical potential on the center of mass energy in nucleus-nucleus 
collisions can be parametrized by \cite{cleymans} 
\begin {equation}
T(\mu_B) = a - b\mu_B^2 - c\mu_B^4
\end {equation}
where $a = (0.166 \pm 0.002) $ $GeV$, $b = (0.139 \pm 0.016) $ ${ GeV^{-1}}$,  
$c = (0.053 \pm 0.021) $ $GeV^{-3} $ and
\begin {equation}
\mu_B (\sqrt s_{NN}) = d/{(1+ e\sqrt s_{NN})}
\end {equation}
with $d$, $e$ given in {Table 1} in \cite{karsch-strange}.
The ratio of baryon to strangeness chemical potential on the freeze-out 
curve shows a weak dependence on the collision energy
\begin{equation}
{\mu_S\over\mu_B} \sim 0.164 + 0.018 \sqrt s_{NN}
\end{equation}
  
The pressure of the strongly interacting matter can be written as,
\begin {equation}
P(T,\mu_B,\mu_Q,\mu_S)=-\Omega (T,\mu_B,\mu_Q,\mu_S),
\label{pres}
\end {equation}
where $T$ is the temperature, $\mu_B$ is the baryon (B) chemical potential, 
$\mu_Q$ is the charge (Q) chemical potential and $\mu_S$ is the 
strangeness (S) chemical potential. From the usual thermodynamic
relations the first derivative of pressure with respect to
quark chemical potential $\mu_q$ is the quark number density and
the second derivative corresponds to the quark number susceptibility (QNS).

 Minimizing the thermodynamic potential numerically with
respect to the fields $\sigma_u$, $\sigma_d$, $\sigma_s$, $\Phi$ and 
$\bar \Phi$, the mean field value for pressure can be obtained 
using the equation (\ref{pres}) \cite {deb}.
The scaled pressure obtained in a given range of chemical potential 
at a particular temperature can be expressed in a Taylor series as,
\begin {equation}
\frac{p(T,\mu_B,\mu_Q,\mu_S)}{T^4}=\sum_{n=i+j+k}c_{i,j,k}^{B,Q,S}(T) 
           (\frac{\mu_B}{T})^i (\frac{\mu_Q}{T})^j (\frac{\mu_S}{ T})^k
\end{equation}
where,
\begin {equation}
c_{i,j,k}^{B,Q,S}(T)={\frac{1}{i! j! k!} 
\frac{\partial^i}{\partial (\frac{\mu_B}{T})^i} 
\frac{\partial^j}{\partial (\frac{\mu_Q}{T})^j} 
\frac{\partial^k {(P/T^4)}}{\partial (\frac{\mu_S}{T})^k}}\Big|_{\mu_{q,Q,S}=0}
\end{equation}
where $\mu_B$, $\mu_Q$, $\mu_S$ are related to the flavor chemical potentials  
$\mu_u$, $\mu_d$, $\mu_s$ as,  
\begin {equation}
  \mu_u=\frac{1}{3}\mu_B+\frac{2}{3}\mu_Q,~~~ 
  \mu_d=\frac{1}{3}\mu_B-\frac{1}{3}\mu_Q,~~~
  \mu_s=\frac{1}{3}\mu_B-\frac{1}{3}\mu_Q-\mu_S
\label{mureln1}
\end {equation}
In this work we evaluate the correlation coefficients up to fourth order
which are generically given by;
\begin{equation}
c_{i,j}^{X,Y}=\dfrac{1}{i! j!}\dfrac{\partial^{i+j}\left(P/T^4\right)}
{{\partial\left({\frac{\mu_X}{T}}\right)^i}{\partial\left({\frac{\mu_Y}{T}}
\right)^j}}
\end{equation}
where, X and Y each stands for B, Q and S with $X\neq Y$.
To extract the Taylor coefficients, first the pressure is obtained as 
a function of different combinations of chemical potentials for each 
value of T and fitted to a polynomial about zero chemical potential
using the gnu-plot fit program \cite{gnu}. Stability of the fit has 
been checked by varying the ranges of fit and simultaneously keeping 
the values of least squares to $10^{-10}$ or even less. At low temperature
fluctuations of a particular charge are dominated by lightest hadrons carrying
that charge. The dominant contribution to $\chi^2_B$ at low temperatures 
comes from protons (lightest baryon), while $\chi^2_S$ receives leading 
contribution from kaons (lightest strange hadron) and $\chi^2_Q$ from 
pions (lightest charged hadron). Since, pion is lighter than proton and kaon,
magnitude of $\chi^2_Q$ is more than that of $\chi^2_B$ and $\chi^2_S$.

\vskip 0.2in
{\subsection{Results}}
The experimental results for volume independent cumulant ratios of net-kaon
distributions are presented
for the first time for all BES energies ${\sqrt s_{NN}}= 7.7, 11.5, 14.5, 
19.6, 27, 39, 62.4$ and $200 \text{GeV}$ for top central and peripheral 
collisions. We have presented our results for four set of parameters of PNJL 
model with six-quark and eight-quark interactions. Also we have compared our 
results with the recent experimental result and HRG model results 
\cite{karsch-strange}. The ratios of charge fluctuations for different moments 
have been considered as they are 
independent of definitions of the interaction volume and also are more 
sensitive to produce correlation length.  
%______________________________________________________________________
\begin{figure}[htb]
\centering
\includegraphics[scale=0.3]{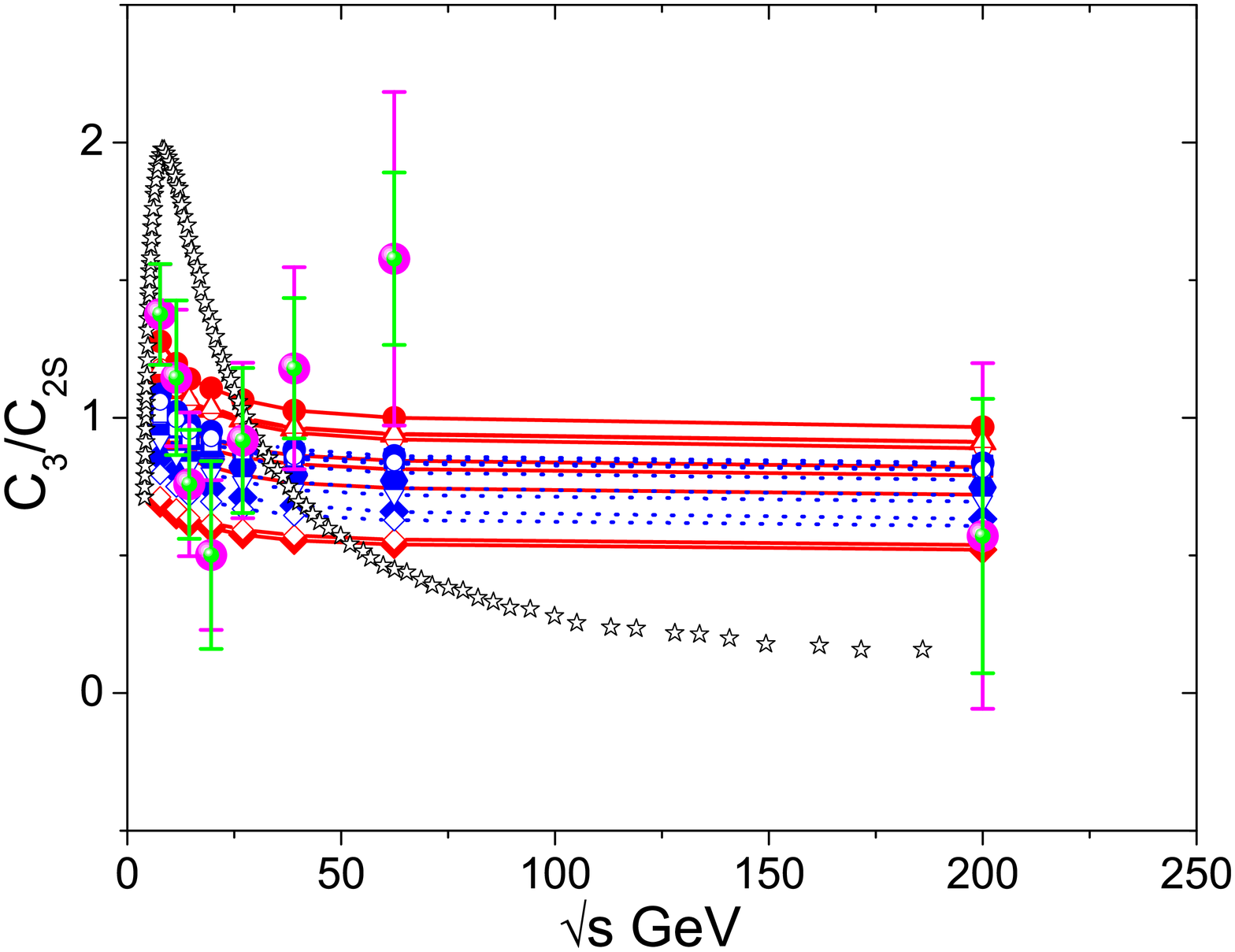}
\includegraphics[scale=0.3]{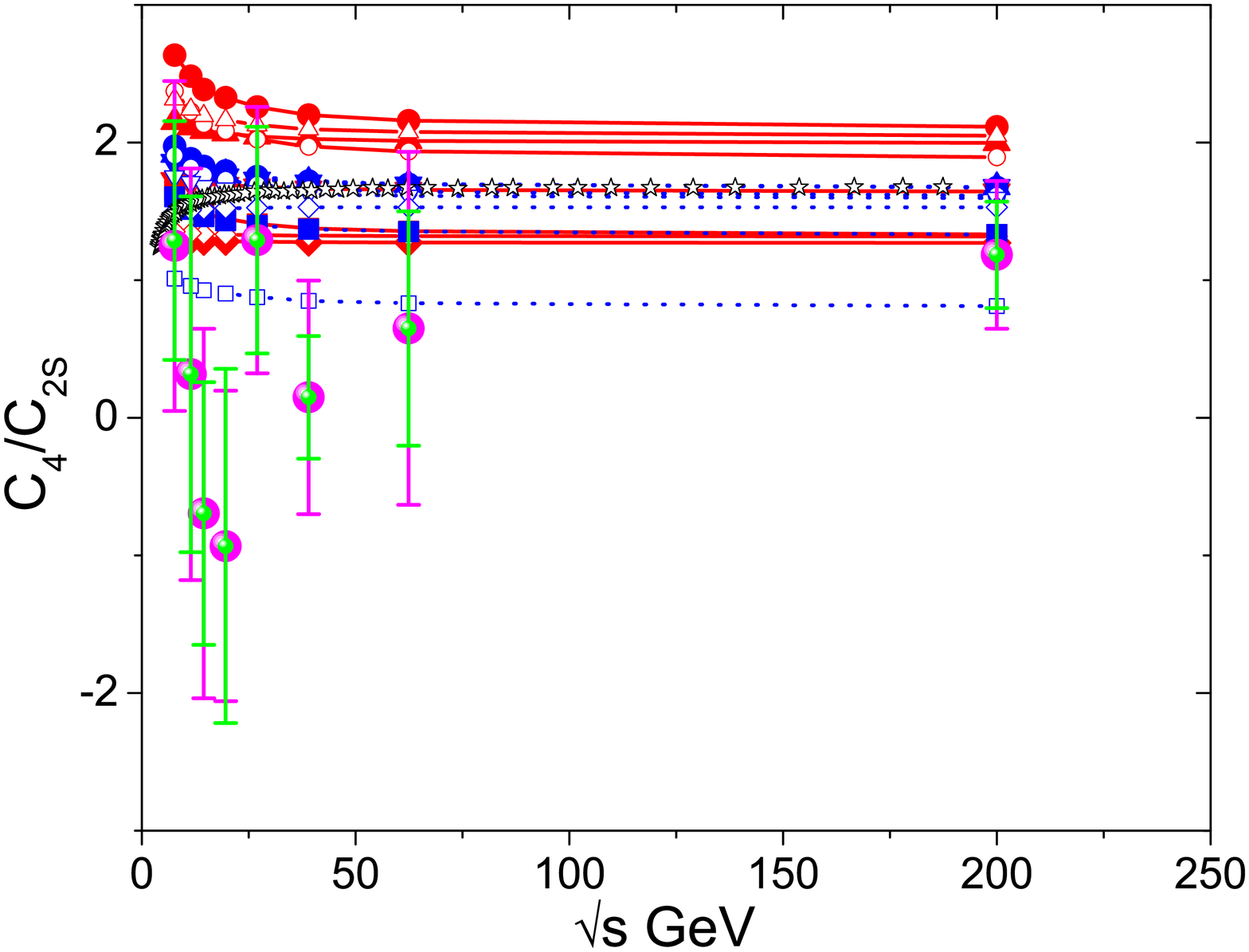}
\caption{(Color online) Kurtosis (right panel) and skewness (left panel) 
of strangeness fluctuations for 
different PNJL parameter sets and comparison with recent STAR data and 
HRG model data. PNJL 6 quark data are plotted with closed symbols $\bullet$ and 8 quark
data are plotted with open symbols $\circ$. $R=2 fm$ data are denoted by straight line
in all red symbols ${\red {-}}$ and $R=4 fm$ data are denoted by dotted lines 
in all blue symbols ${\blue {--}}$. The temperature scheme for different plots 
are as follows : 
$T=100 MeV$ as square $\square$, $T=130 MeV$ as circle $\circ$, $T=150 MeV$ as up triangle $\vartriangle$, 
$T=170 MeV$ as down triangle $\triangledown$ and $T=200 MeV$ as rhombus $\Diamond$. Black star ${\black \star}$ is denoted
as HRG data. Green circles ${\green \circ} $describe 70-80 percent peripheral collision and 
magenta circles ${\magenta \circ}$ are denoted as 0 - 5 percent collision 
in recent STAR preliminary result.} 
\label {c3c4}
\end{figure} 

Figure ({\ref {c3c4}}) shows the variation of 
$C_3 / C_{2_S}$ and $C_4 / C_{2_S}$ fluctuation with respect to 
different collision energies for
PNJL model with four-quark and six-quark interactions with finite volume
system with $R=2fm$ and $R=4fm$. 
As we increase the temperature $C_3 / C_{2_S}$ 
decreases quantitatively. Also for higher collision energy it decreases
for each temperature. $C_4 / C_{2_S}$ has similar features as 
$C_3 / C_{2_S}$. The values of $C_4 / C_{2_S}$ becomes higher for 
smaller collision energies and gradually decreases with increasing energy.
For higher temperature the value decreases quantitatively. 
We have compared our results with the recent experimental data from STAR and 
with the Hadron Resonance gas model (HRG) results. In recent experimental data
no significant deviation has been found with respect to the
Poisson expectation value within statistical and systematic
uncertainties for both the moments {\cite {thader}}.
For the skewness ratio our model results 
for a particular temperature $T=130 MeV$ are very near to the Poisson 
expectation value. The results from the PNJL model are in good agreement
with the experimental results.  
For the collision energy $ \sqrt s <  27 \text{ GeV}$, there is 
an enhancement of fluctuation for PNJL model. Also in case of STAR results, 
there is an deviation from Poisson expectation value.  
Although the results for both skewness and kurtosis have 
qualitative similarities for 
both PNJL and HRG model, the values have quantitative differences.

We now set out to present the results for correlations among different 
conserved charges. In QGP, as baryon number as well as electric charge are
carried by different flavors of quarks, a strong correlation is expected 
between B-Q, Q-S as well as B-S correlations. Also it is expected that the 
heavier particle will interact with the sigma field more strongly than 
the lighter particle. So it is important to study the different freeze-out stages of the produced QGP medium.
On the other hand, in the hadronic sector
presence of baryons and mesons would generate an entirely different type 
of correlations between these quantities. Hence these correlations are expected 
to show changes across the freeze-out which are characteristics of the
changes in the relevant degrees of freedom. 

 Let us consider the baryon-strangeness (BS) correlation. In 
figure (\ref {BS11B2}) the leading order BS correlation is shown for 4 sets of
PNJL model. The correlation normalized to the baryon number fluctuations are 
given by
\begin{eqnarray*}
\textrm{C}_\textrm{BS} = - \dfrac{\chi_{BS}}{\chi_{SS}} = -\frac{1}{2} 
\dfrac{c_{11}^{BS}}{c_2^S} \\
\textrm{C}_\textrm{SB} = - \dfrac{\chi_{BS}}{\chi_{BB}} = -\frac{1}{2} 
\dfrac{c_{11}^{BS}}{c_2^B}
\end{eqnarray*}
where we have used the notation; 
$\chi_{XY}=\dfrac{\partial^2P}{\partial\mu_X\partial\mu_Y}$ and
$\chi_{XX}=\dfrac{\partial^2P}{\partial\mu_X^2}$
Since $C_{BS}$ has entirely different behavior in the hadron gas and in QGP, it
can be a reasonable diagnostic tool for identifying the nature of matter formed
in heavy-ion collisions. 
Figure {\ref {BS11B2}} represents the baryon and strangeness correlation 
normalized to the fluctuation of baryon number for all 4 parameter sets of PNJL
model. As the temperature increases, the value of the ratio $C_{SB}$ decreases.
For smaller collision energy, the value of the fluctuation is large compared 
to the higher collision energy. With the increase in collision energy, 
the value of the baryon chemical potential decreases. So the baryon fluctuation
decreases with 
the decrease of baryon chemical potential. Thus the baryon number and the 
strangeness correlation is much larger than the baryon fluctuations.  
For all parameter 
sets of PNJL model the plots show similar enhancement of fluctuation at
lower collision energy.       
\begin{figure}[htb]
\centering
\includegraphics[scale=0.3]{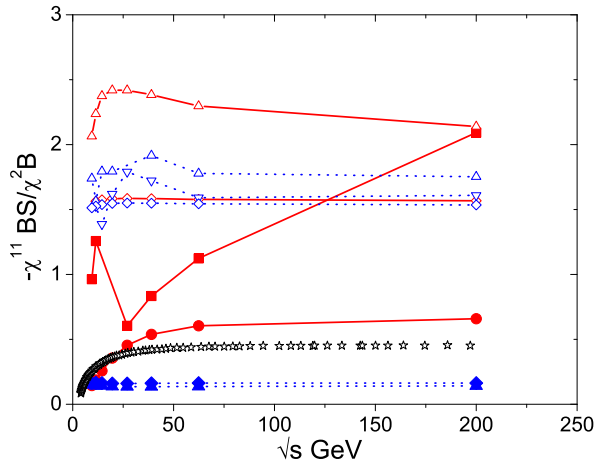}
\caption{(Color online) $\chi^{11}_{BS} / \chi^2_B$ correlations for different 
PNJL parameter sets.  PNJL 6 quark data are plotted with closed symbols $\bullet$ and 
8 quark data are plotted with open symbols $\circ$. 
 $R=2 fm$ data are denoted by straight line
in all red symbols ${\red {-}}$ and $R=4 fm$ data are denoted by dotted lines 
in all blue symbols ${\blue {--}}$. The temperature scheme for different plots 
are as follows : 
$T=100 MeV$ as square $\square$, $T=130 MeV$ as circle $\circ$, $T=150 MeV$ as up triangle $\vartriangle$, 
$T=170 MeV$ as down triangle $\triangledown$ and $T=200 MeV$ as rhombus $\Diamond$. Black star ${\black \star}$ is denoted
as HRG data.}
\label {BS11B2}
\end{figure}

Now we show the behavior of some fourth order correlations - $\chi_{13}^{BS}$,
$\chi_{31}^{BS}$. We have plotted the correlations for 
the four sets of PNJL parameters for different temperatures.  
Figure {\ref {BS13}} represents the $-\chi^{13}_{BS} / \chi^2_B$ and 
$\chi^{31}_{BS}$ correlations. The value
of  $-\chi^{13}_{BS} / \chi^2_B$ is higher for lower collision 
energy for 6q PNJL model. For PNJL model with 8q interaction, correlation 
has qualitative similarity as HRG model. At low collision energy the value
is low compared to the higher collision energy region. But they have
quantitative difference than HRG data. $\chi^{31}_{BS}$ value increases with
increasing temperature and for higher temperature the values are lower for
low collision energy region.   
  
\begin{figure}[htb]
\centering
\includegraphics[scale=0.3]{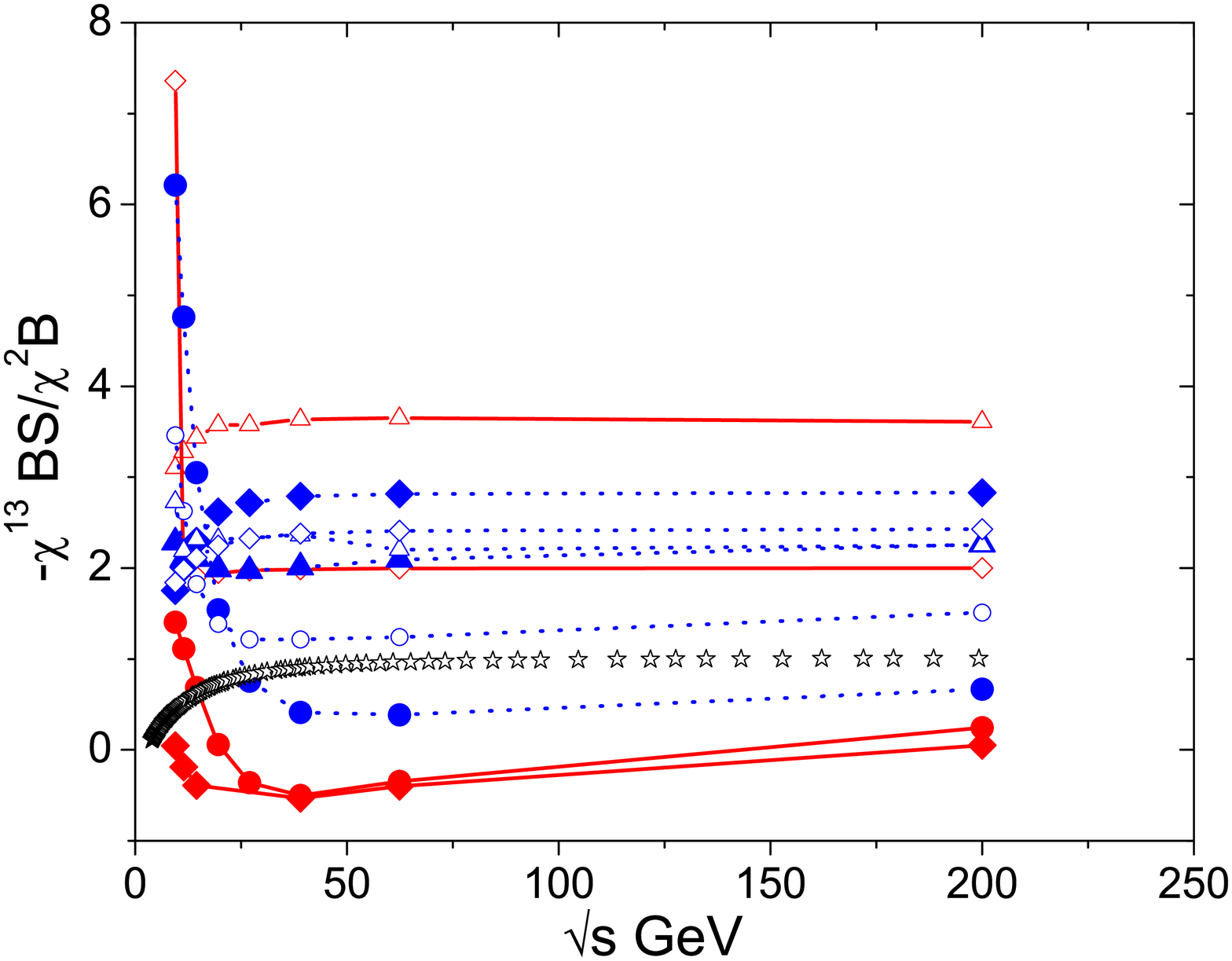}
\includegraphics[scale=0.3]{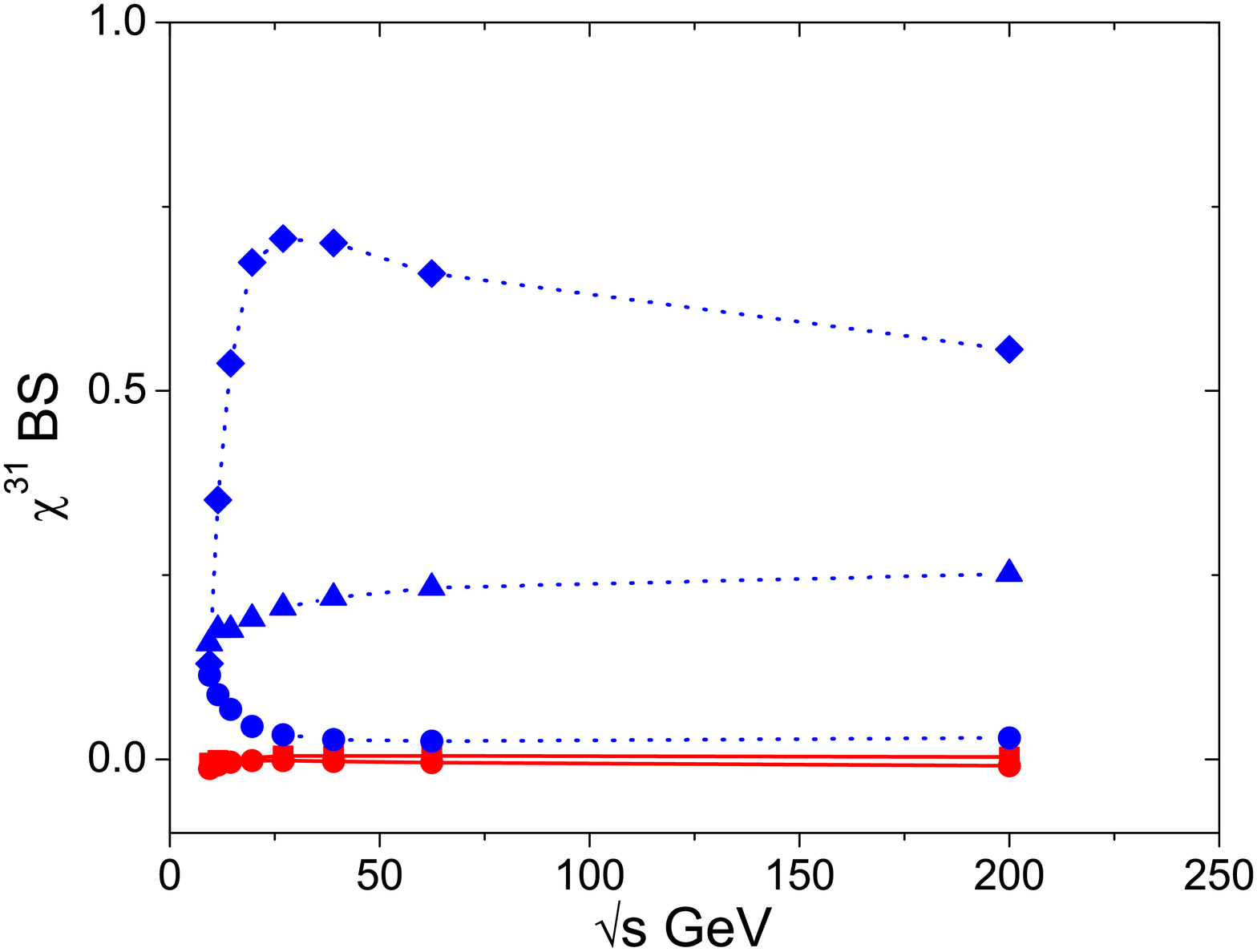}
\caption{(Color online) $\chi^{13}_{BS}$ (left panel) and $\chi^{31}_{BS}$ 
(right panel) correlations for different PNJL parameter sets.
PNJL 6 quark data are plotted with closed symbols $\bullet$ and 8 quark
data are plotted with open symbols $\circ$. 
 $R=2 fm$ data are denoted by straight line
in all red symbols ${\red {-}}$ and $R=4 fm$ data are denoted by dotted lines 
in all blue symbols ${\blue {--}}$. The temperature scheme for different plots 
are as follows : 
$T=100 MeV$ as square $\square$, $T=130 MeV$ as  circle $\circ$, $T=150 MeV$ as up triangle $\vartriangle$, 
$T=170 MeV$ as down triangle $\triangledown$ and $T=200 MeV$ as rhombus $\Diamond$. Black star ${\black \star}$ is denoted
as HRG data. }
\label {BS13}
\end{figure}

 We now turn to baryon-charge (BQ) correlation. In case of electric charge,
fluctuations multiple charged hadrons have larger contribution in higher
moments which results in characteristic deviations of the kurtosis and 
skewness. In figure {\ref {BQ11B2}} the leading order baryon-charge correlation 
has been normalized by $\chi^2_B$. For lower collision energy there is an 
enhanced fluctuation for all sets of PNJL model. The fluctuation increases
with increasing temperature which indicates the transition region
and also it is more for smaller volume system.     

\begin{figure}[htb]
\centering
\includegraphics[scale=0.3]{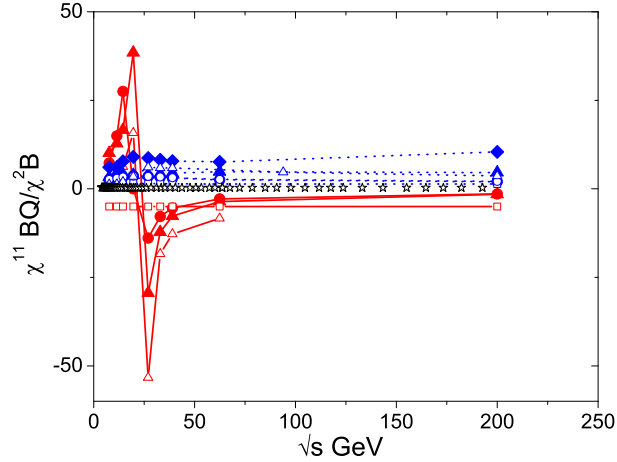}
\caption{(Color online) $\chi^{11}_{BQ} / \chi^2_B$ correlations for different PNJL parameter sets.
PNJL 6 quark data are plotted with closed symbols $\bullet$ and 8 quark
data are plotted with open symbols $\circ$.
 $R=2 fm$ data are denoted by straight line
in all red symbols ${\red {-}}$ and $R=4 fm$ data are denoted by dotted lines 
in all blue symbols ${\blue {--}}$. The temperature scheme for different plots 
are as follows : 
$T=100 MeV$ as square $\square$, $T=130 MeV$ as  circle $\circ$, $T=150 MeV$ as up triangle $\vartriangle$, 
$T=170 MeV$ as down triangle $\triangledown$ and $T=200 MeV$ as rhombus $\Diamond$. Black star ${\black \star}$ is denoted
as HRG data. }
\label {BQ11B2}
\end{figure}

In fig. \ref{BQ13} the fluctuation increases for lower collision energy for 
system with $R=2 fm$. The value of the higher order correlation increases 
quantitatively with temperature. But for the finite volume system with 
$R=4 fm$, the situation 
is different. The value of $\chi^{13}_{BQ}$ correlation is higher near the 
transition temperature, but at lower collision energy the value decreases.
In fig. \ref{BQ13} there is an enhanced fluctuation at lower collision energy 
near the transition temperature. The value of the fluctuation is low
quantitatively for higher temperature.
\begin{figure}[htb]
\centering
\includegraphics[scale=0.3]{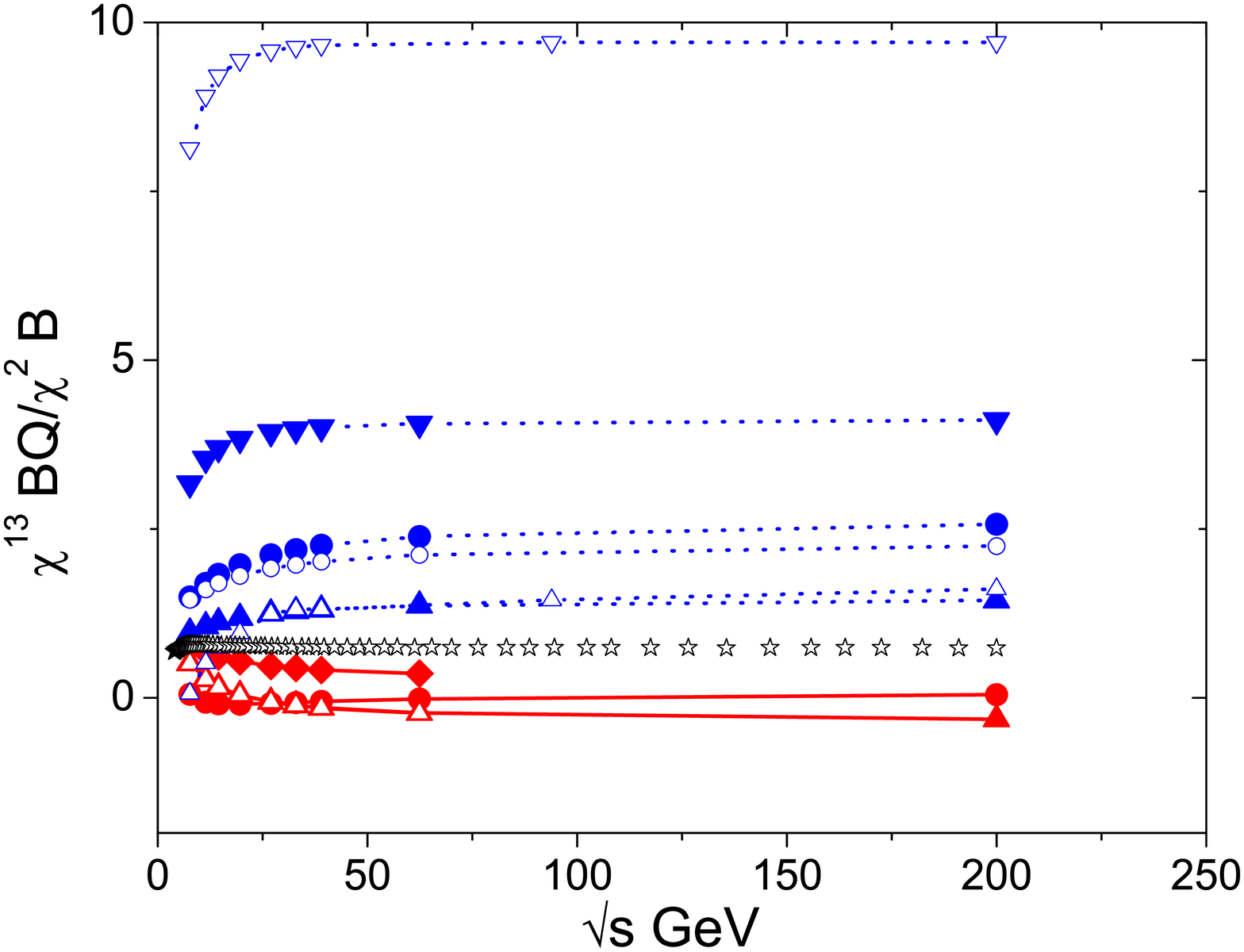}
\includegraphics[scale=0.3]{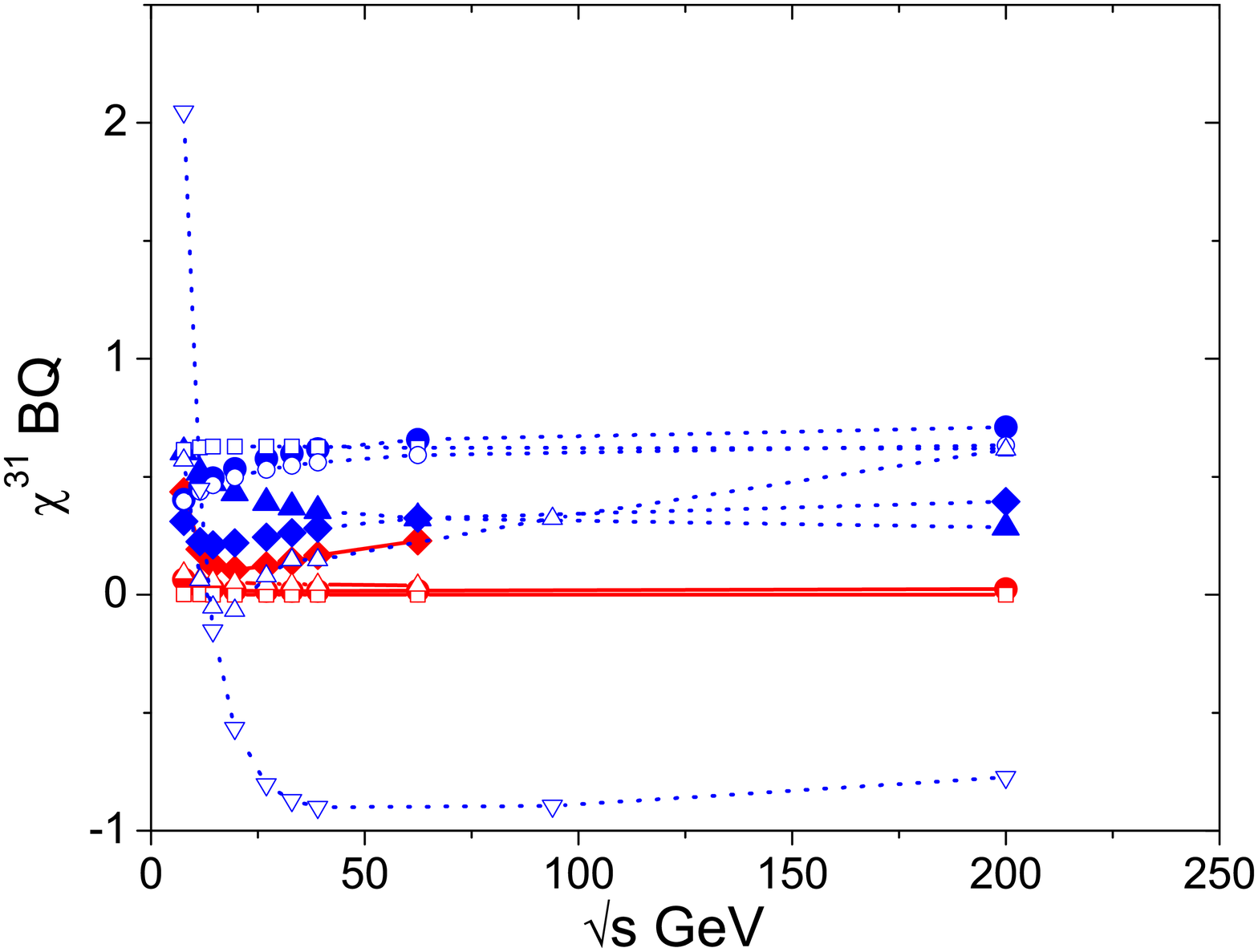}
\caption{(Color online) $\chi^{13}_{BQ}$ (left panel) and $\chi^{31}_{BQ}$
(right panel) correlations for different PNJL parameter sets.
PNJL 6 quark data are plotted with closed symbols $\bullet$ and 8 quark
data are plotted with open symbols $\circ$. 
 $R=2 fm$ data are denoted by straight line
in all red symbols ${\red {-}}$ and $R=4 fm$ data are denoted by dotted lines 
in all blue symbols ${\blue {--}}$. The temperature scheme for different plots 
are as follows : 
$T=100 MeV$ as square $\square$, $T=130 MeV$ as  circle $\circ$, $T=150 MeV$ as up triangle $\vartriangle$, 
$T=170 MeV$ as down triangle $\triangledown$ and $T=200 MeV$ as rhombus $\Diamond$. Black star ${\black \star}$ is denoted
as HRG data. }
\label {BQ13}
\end{figure}

Now we will discuss the leading order and higher order charge-strangeness 
correlations.  As in the case of the baryon number, the charge is also 
strongly correlated to strangeness through strange quarks. At lower collision 
energy and near transition temperature 
there is an enhanced fluctuation at all temperature in fig. \ref {QS11}. 
The value of QS correlation increases quantitatively with the temperature.   
$\chi^{13}_{BS}$ increases as we increase the temperature and for lower 
collision energy. They have similar behavior as for the BS correlations.
Therefore the BS and QS correlations can be used complimentary to 
understand the state of affairs in heavy-ion collisions. 

\begin{figure}[htb]
\centering
\includegraphics[scale=0.3]{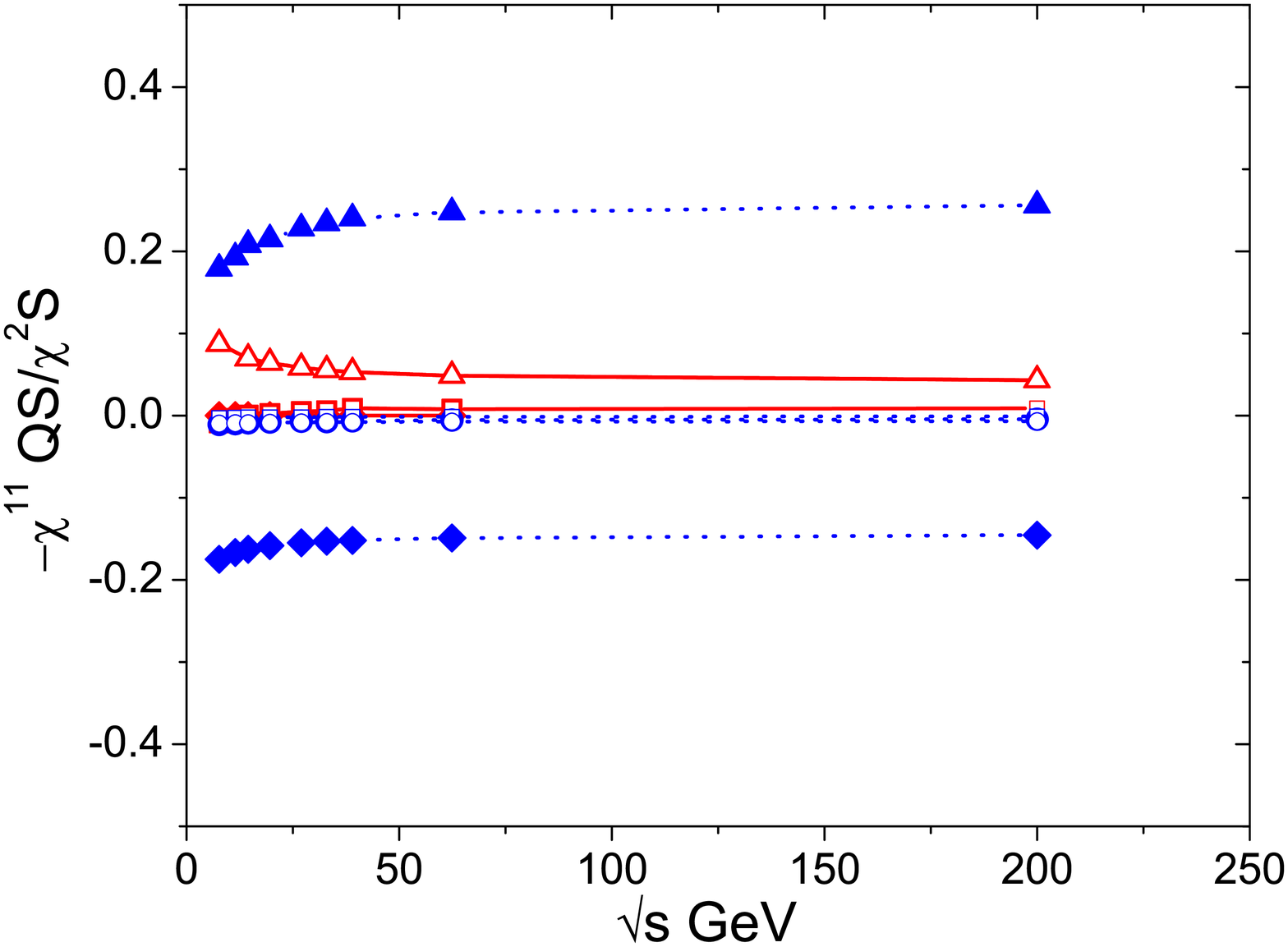}
\caption{(Color online)$\chi^{11}_{QS}$ correlations for different PNJL parameter sets.
PNJL 6 quark data are plotted with closed symbols $\bullet$ and 8 quark
data are plotted with open symbols $\circ$. 
 $R=2 fm$ data are denoted by straight line
in all red symbols ${\red {-}}$ and $R=4 fm$ data are denoted by dotted lines 
in all blue symbols ${\blue {--}}$. The temperature scheme for different plots 
are as follows : 
$T=100 MeV$ as square $\square$, $T=130 MeV$ as  circle $\circ$, $T=150 MeV$ as up triangle $\vartriangle$, 
$T=170 MeV$ as down triangle $\triangledown$ and $T=200 MeV$ as rhombus $\Diamond$.}
\label {QS11}
\end{figure}
  
\begin{figure}[htb]
\centering
\includegraphics[scale=0.3]{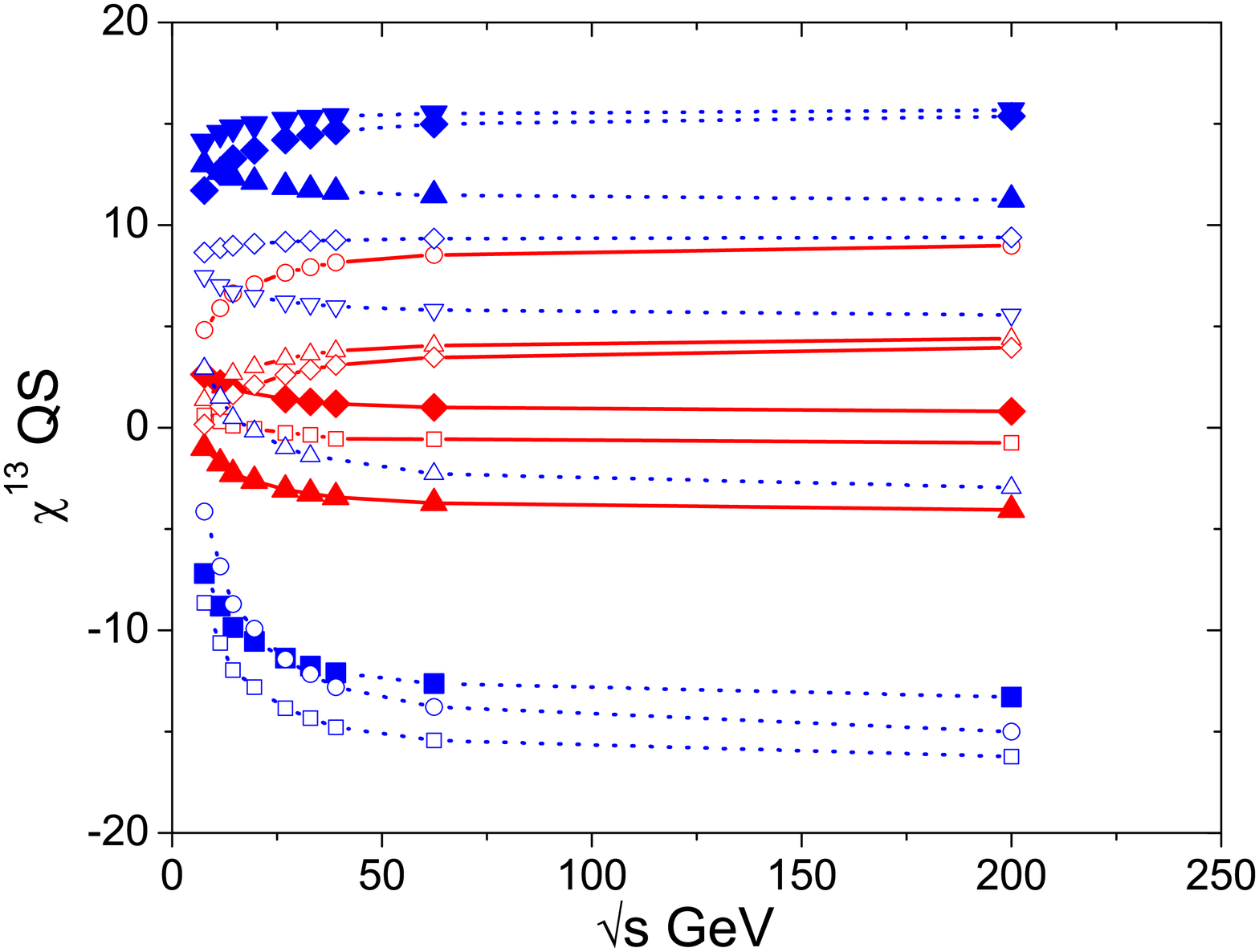}
\includegraphics[scale=0.3]{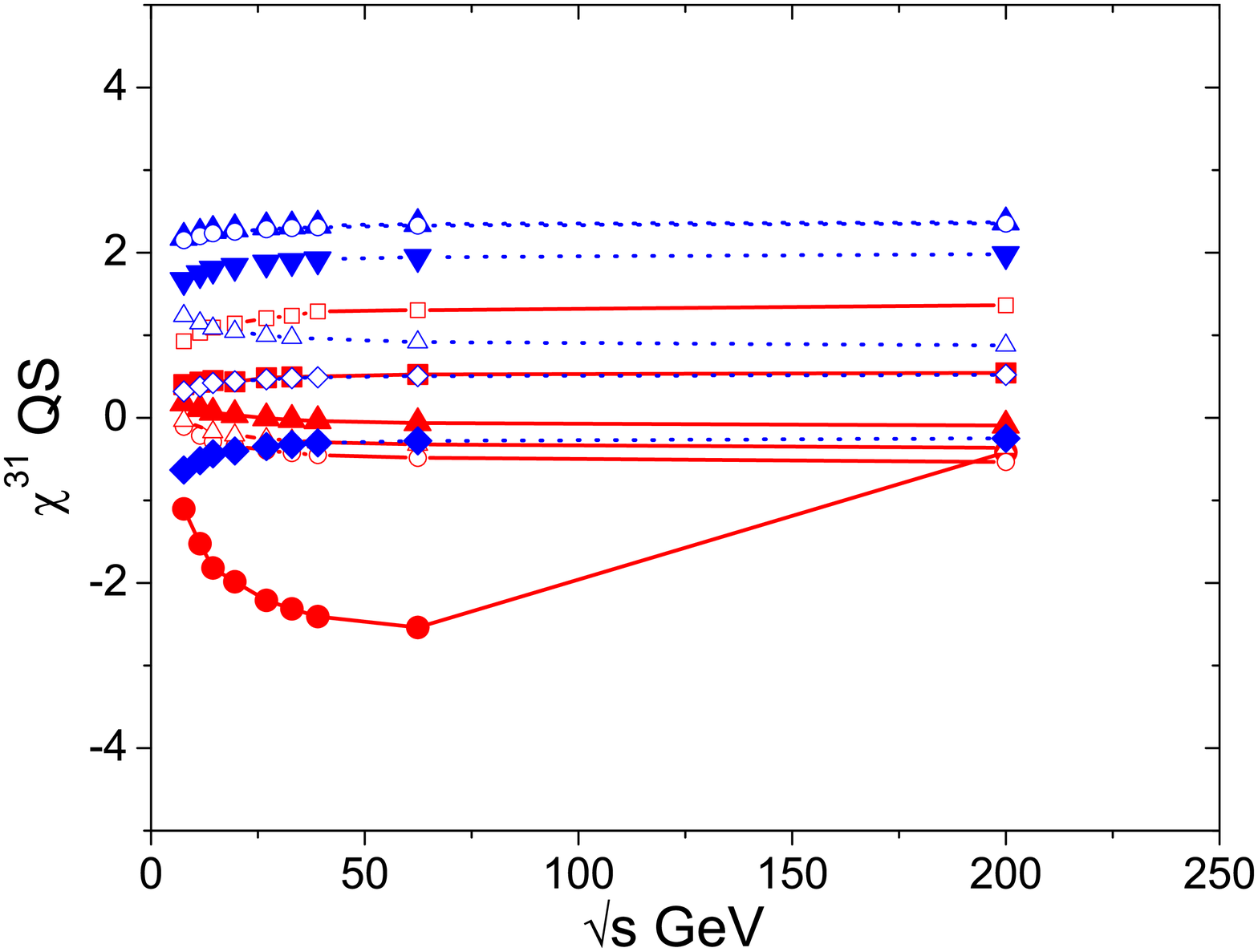}
\caption{(Color online) $\chi^{13}_{QS}$ (left panel) and $\chi^{31}_{QS}$
(right panel) correlations for different PNJL parameter sets.
PNJL 6 quark data are plotted with closed symbols $\bullet$ and 8 quark
data are plotted with open symbols $\circ$.
 $R=2 fm$ data are denoted by straight line
in all red symbols ${\red {-}}$ and $R=4 fm$ data are denoted by dotted lines 
in all blue symbols ${\blue {--}}$. The temperature scheme for different plots 
are as follows : 
$T=100 MeV$ as square $\square$, $T=130 MeV$ as  circle $\circ$, $T=150 MeV$ as up triangle $\vartriangle$, 
$T=170 MeV$ as down triangle $\triangledown$ and $T=200 MeV$ as rhombus $\Diamond$.}
\label {QS13}
\end{figure}

 \section{summary}
We have discussed properties of net kaon fluctuations in nuclear matter 
within PNJL model. We have considered the
ratio of fourth order moment to second order moment (kurtosis) and the 
third order moment to the second order moment (skewness) of strangeness 
fluctuations. We have also focused on the cross correlations related to baryon
number, strangeness and electric charge conservation. All the correlations
were obtained by fitting the pressure in a Taylor series expansion around 
the finite baryon, charge and strangeness chemical potentials. The baryon,
charge and strangeness chemical potentials are obtained from the freeze-out
curve which depends on the collision energies in the BES scan at the heavy-ion
collision experiment. The results are shown for PNJL model with 6 quark
and 8 quark interactions and two finite volume systems with lateral size
$R=2 fm$ and $R=4fm$.

  Skewness and kurtosis of strangeness fluctuation in PNJL model have 
similar features along the collision energy of heavy ion experiments. As we
increase the temperature both skewness and kurtosis value decreases 
quantitatively. For collision energy less than $27$ $GeV$, the value of 
kurtosis and skewness are higher. The recent experimental observations 
show no significant
deviation from Poisson expectation value for both the observables. However
there is small deviations for skewness and kurtosis for low collision energy.
Similarly in PNJL model we have found an enhancement of fluctuations for 
low collision energy less than $27 GeV$. Also near the transition temperature
the skewness ratio is very near to the Poisson expectation value.     

 The various correlators have been discussed to understand the matter created
in the heavy ion collision experiments. The leading order coefficients can be
most useful for identifying if the QGP is formed, while the higher order 
coefficients could identify the crossover region. We have noted a qualitative
similarity of the leading order correlators of BS and QS with HRG model
data. However they have a quantitative differences. The $\chi^{11}_{BQ}$
has large fluctuations at lower collision energies which differs from HRG
model data both qualitatively and quantitatively.

 For higher order correlators containing strangeness $\chi_{BS}$ and 
$\chi_{QS}$ show similar behavior near low collision energy region. 
All the higher order cross correlations show increase or decrease of 
fluctuation at low collision energy. This might indicate the location of 
critical region in heavy-ion collision experiment.      

 The study of various equilibrium thermodynamic measurements of the correlators
using PNJL model would be helpful in determining the finite temperature finite 
density behavior of the hadronic sector. comparison of PNJL results with 
the experimental value will ensure the understanding of the physics behind the 
critical region and to locate the critical point in the strongly interacting 
matter.   

\acknowledgements P.D would like to thank Indian Institute of Technology Bombay
for financial support. The part of the work has been published in the 
Few Body System. 59 (2018) no. 4, 55 of the Light Cone Conference 2017.

\end{document}